\documentclass[conference]{IEEEtran}
\IEEEoverridecommandlockouts
\usepackage{orcidlink}

\usepackage{cite}
\usepackage{amsmath,amssymb,amsfonts}
\usepackage{algorithmic}
\usepackage{graphicx}
\usepackage{textcomp}
\usepackage[table]{xcolor}
\def\BibTeX{{\rm B\kern-.05em{\sc i\kern-.025em b}\kern-.08em
    T\kern-.1667em\lower.7ex\hbox{E}\kern-.125emX}}

\usepackage[misc]{ifsym}
\usepackage{booktabs}
\usepackage{multirow}

\usepackage{tikz}
\usetikzlibrary{arrows.meta,positioning,calc,fit,backgrounds,shadows.blur}

\usepackage{hyperref}
\hypersetup{
    colorlinks=true,      
    linkcolor=red,      
    citecolor=blue,       
    urlcolor=blue,       
    filecolor=magenta,    
}

\begin{document}

\title{Unsupervised Data Drift Monitoring for\\Deployed Deepfake Speech Detectors
\thanks{$^*$ Equal contribution. \textsuperscript{\Letter} Corresponding author.}
\thanks{This work was supported by the Brno University of Technology internal project FIT-S-26-9011. Computational resources were provided by the e-INFRA CZ project (ID:90254), supported by the Ministry of Education, Youth and Sports of the Czech Republic. The authors acknowledge using ChatGPT and Grammarly to assist with grammar and paraphrasing.}
}

\author{\IEEEauthorblockN{Lukáš Prokeš\textsuperscript{$*$}, Vojtěch Staněk\,\orcidlink{0009-0000-5722-0571}\,\textsuperscript{$*$, \Letter}, Anton Firc\,\orcidlink{0000-0002-4717-1910}, Kamil Malinka\,\orcidlink{0000-0002-9009-2193}}
\IEEEauthorblockA{\textit{Security@FIT, Brno University of Technology, Brno, Czech Republic} \\
\{xproke14, istanek, ifirc, malinka\}@fit.vut.cz}}

\maketitle


\begin{abstract}
Deepfake speech detectors are typically evaluated as binary classifiers on fixed benchmarks, but deployed systems face evolving attacks, codecs, speakers, and recording conditions, with labels unavailable during operation. This makes it difficult to know when a detector operates in familiar conditions and can be relied on. We propose an unsupervised drift monitor that operates on the internal embeddings of a deployed detector. The monitor compares each input against bona fide and spoof references using regularized Mahalanobis distances. These distances define trusted, ambiguous, and novel regions of the detector embedding space, allowing unfamiliar inputs to be routed for review. We evaluate the monitor on four modern detectors on three datasets. The monitor raises alerts for dataset mismatches and substantially reduces the security-critical case of accepting deepfakes at the cost of reduced automatic coverage. The results also show that drift affects both classes: bona fide recordings can move away from the training reference, increasing alert rates even in in-domain settings. Our findings suggest that deployed deepfake speech detectors should be treated not only as a binary classification, but also as a reliability problem.
\end{abstract}


\section{Introduction}

Voice cloning and speech synthesis systems have advanced to the point where convincing synthetic speech can be generated even from limited speaker material~\cite{firc2025evaluation}. This progress has intensified the need for automatic speech deepfake detectors~\cite{wang24_asvspoof}, especially in forensic, journalistic, biometric, and platform-moderation scenarios where manipulated audio may be used to impersonate a speaker or alter public perception~\cite{Malinka2024Comprehensive}. Recent detectors based on large self-supervised learning (SSL) models achieve very low error rates on established benchmarks~\cite{chen24_asvspoof}, but their deployment reliability remains fragile~\cite{muller2022inthewild}. A detector is typically trained on a limited set of spoofing attacks, whereas real deployments face an open, evolving stream of synthesis methods, codecs, speakers, channels, and post-processing conditions~\cite{tsutsumi26_odyssey}.

The unseen nature of operating conditions creates a practical failure mode: the detector may continue to output confident, but wrong decisions when its operating domain has shifted~\cite{wang2025datadrift}. Standard benchmark evaluation can reveal this degradation only after labels are available~\cite{Liu2023}, but \textbf{labels are not available during operation on real-world data!} The relevant question is not only whether a recording is bona fide or deepfake, but also whether the detector is still operating in familiar conditions where its decision can be trusted. A deepfake speech detector that fails silently without any warning is operationally very dangerous, even if it performs well on the benchmark for which it was trained.

In this paper, we study whether data drift, i.e., a mismatch between the detector's training and deployment recordings, can be detected directly from the detector's internal representation, without retraining the detector and without requiring labels for incoming data~\cite{wang2022dataselection, furuhashi2024dataselection}. We treat the detector as a fixed component and add a monitoring layer on top of its utterance embedding, similar to Wang et al.~\cite{wang2025datadrift}. We propose a monitoring framework using the Mahalanobis distance~\cite{mahalanobis_neurips} to compare each embedding against two reference profiles constructed from the detector's training data: a bona fide reference (known genuine speech) and a deepfake reference (spoofs seen during training). Distances to these references are then used to determine whether a recording resembles known bona fide speech, resembles known attacks, falls close to both references, or lies far from both.

This two-reference formulation changes the role of drift monitoring. Instead of only asking whether the incoming data distribution has shifted~\cite{wang2025datadrift}, the proposed monitor provides an interpretable categorization of the drift. This distinction is important for deployment because not all drift has the same security meaning. In particular, an attack that lies far from bona fide speech may be easy for the detector, whereas an attack whose embeddings move toward the bona fide region may be more dangerous. Similarly, genuine speech may drift due to channel, codec, language, or speaker differences, thereby increasing false alarms even in the absence of a new spoofing method. This joint view turns embedding drift into an operational reliability map, where each input is interpreted not only by its predicted class, but also by whether it lies in a trusted, ambiguous, or novel region of the detector's embedding space.

We evaluate the proposed drift monitor on two detector architectures: AASIST~\cite{Jung2022aasist} and MHFA~\cite{peng2022} with \mbox{XLS-R}~\cite{Babu2021} frontend, each trained either on ASVspoof~2019~LA\cite{todisco2019asvspoof} (AS19) or ASVspoof~5~\cite{wang24_asvspoof} (AS5); four detection systems in total.
Each detector is evaluated on the AS19 and AS5 eval subsets and the In-the-Wild (ITW)~\cite{muller2022inthewild} dataset, which provides a more realistic open-set setting. This setup allows us to test whether the monitor captures operational drift across varying data and deployment conditions. The results show that the monitor correctly alerts dataset mismatch, substantially reduces deepfake false acceptance, but also exposes bona fide drift as an important source of deployment uncertainty. 

Our findings suggest that deepfake speech detection should be treated not only as a binary classification problem, but also as a reliability problem. In open-set deployment, a detector should not only decide whether an input is genuine or fake, but should also indicate uncertainty when its own decision falls outside the conditions under which it was trained.

The contributions of this paper are threefold:
\begin{itemize}
\item We propose an \textbf{unsupervised drift monitor} for deployed deepfake speech detectors that operates on utterance embeddings and requires neither retraining the detector nor labels for incoming recordings.
\item We introduce an \textbf{operational two-reference drift categorization} that compares each recording against bona fide and known-attack references, allowing the monitor to distinguish between familiar, ambiguous, and novel data.
\item We experimentally show that the \textbf{monitor reduces deepfake false acceptance} and reveals bona fide drift as an important deployment factor.
\end{itemize}

\section{Background}


Deepfake speech detection is commonly formulated as binary classification between bona fide and spoofed speech~\cite{wang24_asvspoof}. Recent detectors often combine large SSL models such as XLS-R~\cite{Babu2021} with pooling and classification backends, such as AASIST~\cite{borodin24_asvspoof} or MHFA~\cite{rohdin24_asvspoof}. Their performance is typically measured by Equal Error Rate or detection cost~\cite{wang24_asvspoof} on benchmark datasets such as the ASVspoof series~\cite{todisco2019asvspoof}. These benchmarks have been central to progress in spoofing countermeasures because they provide controlled protocols, common evaluation data, and increasingly diverse spoofing conditions. However, deployment differs from benchmark evaluation in a crucial way. A detector is trained on a finite set of known attacks, while future inputs may be produced by newer synthesis models, vocoders, codecs, or post-processing pipelines that were not available during training. The deployment problem is therefore open-set: the detector must process recordings that may lie outside the distribution on which it was trained and calibrated. In this setting, high benchmark accuracy does not by itself guarantee reliable operation after deployment~\cite{muller2022inthewild}.


This mismatch can be viewed as \textit{data drift} or out-of-domain operation. The labels remain the same, bona fide or spoof, but the input distributions may change~\cite{wang2025datadrift}. This is especially problematic because neural detectors can continue to output confident, but wrong scores, even when inputs no longer correspond to a familiar operating regime. Confidence-based rejection may therefore be insufficient, since a detector can remain confident under a distribution shift, posing a serious security threat. 
For deployed detectors, this creates a need for label-free monitoring that estimates not only what class the detector predicts, but also whether the input resembles the bona fide and spoofing conditions under which the detector was trained. This is important because a detector operates reliably only within such familiar conditions~\cite{muller2022inthewild}.
Drift monitoring for speech deepfake detection remains largely unexplored. Wang et al.~\cite{wang2025datadrift} study drift at the dataset level, while Salvi et al.~\cite{salvi2023reliability} estimate sample-level reliability with a learned reliability estimator. Their objective is to predict whether the detector's current decision is \textit{likely} to be correct. Our objective is different: we assess whether an incoming sample remains consistent with the detector's known bona fide and spoof training distributions, regardless of whether the current prediction is correct. This allows distribution shift to be identified even before it results in detector errors and avoids introducing an additional learned reliability predictor.
Our work, therefore, monitors a fixed, i.e., \textit{deployed}, detector: instead of replacing or retraining the classifier~\cite{wang2022dataselection, furuhashi2024dataselection}, we use detector embeddings and training reference distributions to estimate when an automatic decision falls outside the detector's trusted operating regime.

\section{Unsupervised Drift Monitoring}

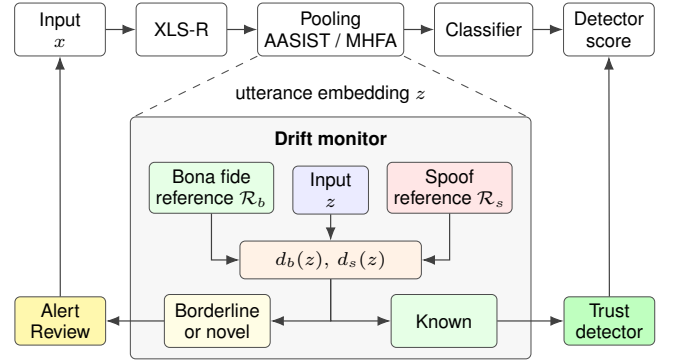
\begin{figure}[t]
\centering
\begin{tikzpicture}[
    font=\sffamily\scriptsize,
    >=Latex,
    node distance=4mm and 4mm,
    every node/.style={font=\sffamily\scriptsize},
    box/.style={
        draw=black!70,
        rounded corners=2pt,
        very thin,
        align=center,
        minimum height=7mm,
        inner sep=2.5pt,
        fill=white
    },
    stage/.style={
        box,
        minimum width=16mm
    },
    smallbox/.style={
        box,
        minimum width=13mm,
        minimum height=5mm
    },
    monbox/.style={
        box,
        fill=gray!6
    },
    title/.style={
        font=\sffamily\scriptsize\bfseries
    },
    arr/.style={
        ->,
        line width=0.45pt,
        draw=black!75
    },
    dashedarr/.style={
        dashed,
        line width=0.4pt,
        draw=black!70
    }
]

\node[stage, minimum width=12mm] (input) {Input\\$x$};
\node[stage, right=of input, minimum width=12mm] (xlsr) {XLS-R};
\node[stage, right=of xlsr, minimum width=18mm] (pool) {Pooling\\AASIST / MHFA};
\node[stage, right=of pool, minimum width=13mm] (clf) {Classifier};
\node[stage, right=of clf, minimum width=12mm] (score) {Detector\\score};

\draw[arr] (input) -- (xlsr);
\draw[arr] (xlsr) -- (pool);
\draw[arr] (pool) -- (clf);
\draw[arr] (clf) -- (score);

\coordinate (tap) at ($(pool.south)+(0,-1mm)$);

\node[monbox, below=8mm of pool, minimum width=53mm, minimum height=32mm] (monitor) {};

\node[font=\sffamily\scriptsize\bfseries, anchor=north] at ($(monitor.north)+(0,-0.7mm)$)
{Drift monitor};

\node[smallbox, fill=green!10, anchor=north west] (refb) at ($(monitor.north west)+(2.5mm,-6.0mm)$) {Bona fide\\reference $\mathcal{R}_b$};

\node[smallbox, fill=red!10, anchor=north east] (refs) at ($(monitor.north east)+(-2.5mm,-6.0mm)$) {Spoof\\reference $\mathcal{R}_s$};

\node[smallbox, fill=blue!8, minimum width=10mm, anchor=north] (zin) at ($(monitor.north)+(0,-6.5mm)$) {Input\\$z$};

\node[smallbox, fill=orange!10, minimum width=24mm, below=3.5mm of zin] (dist) {$d_b(z),\, d_s(z)$};


\node[smallbox, fill=green!10, minimum width=14mm, minimum height=7mm, below=2mm of dist, xshift=15mm] (known)
{Known};

\node[smallbox, fill=yellow!12, minimum width=14mm, minimum height=7mm, below=2mm of dist, xshift=-15mm] (novel)
{Borderline\\or novel};



\node[smallbox, fill=green!25, minimum width=12mm, right=8.8mm of known] (trust)
{Trust\\detector};

\node[smallbox, fill=yellow!40, minimum width=12mm, left=7.7mm of novel] (alert)
{Alert\\Review};

\draw[arr] (zin) -- (dist);
\draw[arr] (dist.south) |- (known.west);
\draw[arr] (dist.south) |- (novel.east);
\draw[arr] (known.east) -- (trust.west);
\draw[arr] (novel.west) -- (alert.east);

\draw[arr] (refb.south) |- (dist.west);
\draw[arr] (refs.south) |- (dist.east);

\draw[arr] (trust.north) -- (score.south);
\draw[arr] (alert.north) -- (input.south);

\draw[dashedarr] ($(pool.south west)!0.0!(pool.south east)$) -- ($(monitor.north west)!0.01!(monitor.north east)$);
\draw[dashedarr] ($(pool.south west)!1.0!(pool.south east)$) -- ($(monitor.north west)!0.99!(monitor.north east)$);

\node[font=\sffamily\scriptsize, align=center] at ($(pool.south)+(0,-5mm)$)
{utterance embedding $z$};

\end{tikzpicture}
\caption{Proposed deployment pipeline. A fixed speech deepfake detector processes the input recording, produces an utterance embedding $z$, and an output score. The drift monitor compares $z$ to both bona fide and spoof references using Mahalanobis distance. Inputs assigned to known regions are trusted, while borderline or novel inputs are routed to review.}
\label{fig:monitor_schema}
\vspace{-1.4em}
\end{figure}

We monitor a fixed deepfake speech detector without changing its parameters and without using labels from the evaluation data. Let $f_\theta$ denote a trained detector and let $z=f_\theta^{emb}(x)\in\mathbb{R}^{1024}$ be the utterance-level embedding extracted before the final classification head from input recording $x$ (see \autoref{fig:monitor_schema}).

\subsection{Reference profiles and covariance regularization}
\label{sec:method}

For each trained detector, we build two reference profiles from the dataset's training split used to train the detector. The bona fide reference $\mathcal{R}_b$ contains embeddings of all bona fide training recordings, and the spoof reference $\mathcal{R}_s$ contains embeddings of all spoof training recordings. 


For each reference \(\mathcal{R}_c\), where \(c\in\{b,s\}\), we model the corresponding training embeddings as a Gaussian reference cloud. We estimate its mean vector \(\mu_c\) and a covariance matrix \({\Sigma}_c\). Because detector embeddings can be highly anisotropic, i.e., variance is concentrated in only a few directions, while many others have near-zero variance, the ordinary covariance matrix $\Sigma_c$ may be poorly conditioned\footnote{This effect was especially visible for systems with the AASIST backend (see \autoref{sec:regularization}). Their embeddings occupied only a small number of effective dimensions, so many covariance directions had extremely small variance. Since the Mahalanobis distance weights deviations by the inverse covariance, these directions can produce artificially large distances.}. 
This makes $\Sigma_c^{-1}$ used in the Gaussian model unstable.
We therefore use shrinkage covariance estimation~\cite{ledoit2004regularization} to compute a regularized covariance
\[
\widehat{\Sigma}_c = (1-\alpha)\Sigma_c + \alpha \lambda_c I,
\quad
\lambda_c = \frac{\mathrm{Tr}(\Sigma_c)}{D},
\]
where \(\Sigma_c\) is the empirical covariance of the reference embeddings, $\text{Tr}(\cdot)$ is the trace operator, \(D\) is the embedding dimension, and \(\alpha\) controls the amount of shrinkage toward an isotropic covariance. In the final monitor, we use \(\alpha=0.9\) as it produced the most stable results in preliminary experiments.

Given an incoming recording $x$, the monitor computes its embedding $z$ and scores it against both references by the Mahalanobis distance using the regularized covariance:
\[
d_c(z)=\sqrt{(z-\mu_c)^T \ \widehat{\Sigma}_c^{-1} (z-\mu_c)},\quad c\in\{b,s\}.
\]
This yields two distances: $d_b(z)$, the distance from the bona fide reference, and $d_s(z)$, the distance from the known-spoof reference. Unlike the detector score, these distances do not directly ask whether the sample is classified as bona fide or spoof. Instead, they ask whether the sample lies in a region of the embedding space that resembles the detector's training bona fide or spoof data.

\subsection{Two-reference categorization}

The monitor converts the two distances $d_b(z), d_s(z)$ into operating categories using thresholds calibrated on the training references. For each reference $\mathcal{R}_c$, the threshold $\tau_c$ is set to the 95th percentile of self-distances of the training embeddings from that reference. A sample is therefore considered close to reference $\mathcal{R}_c$ when $d_c(z)\leq \tau_c$, and far when $d_c(z)>\tau_c$.

\begin{figure}[tbp]
\centering
\includegraphics[width=\linewidth]{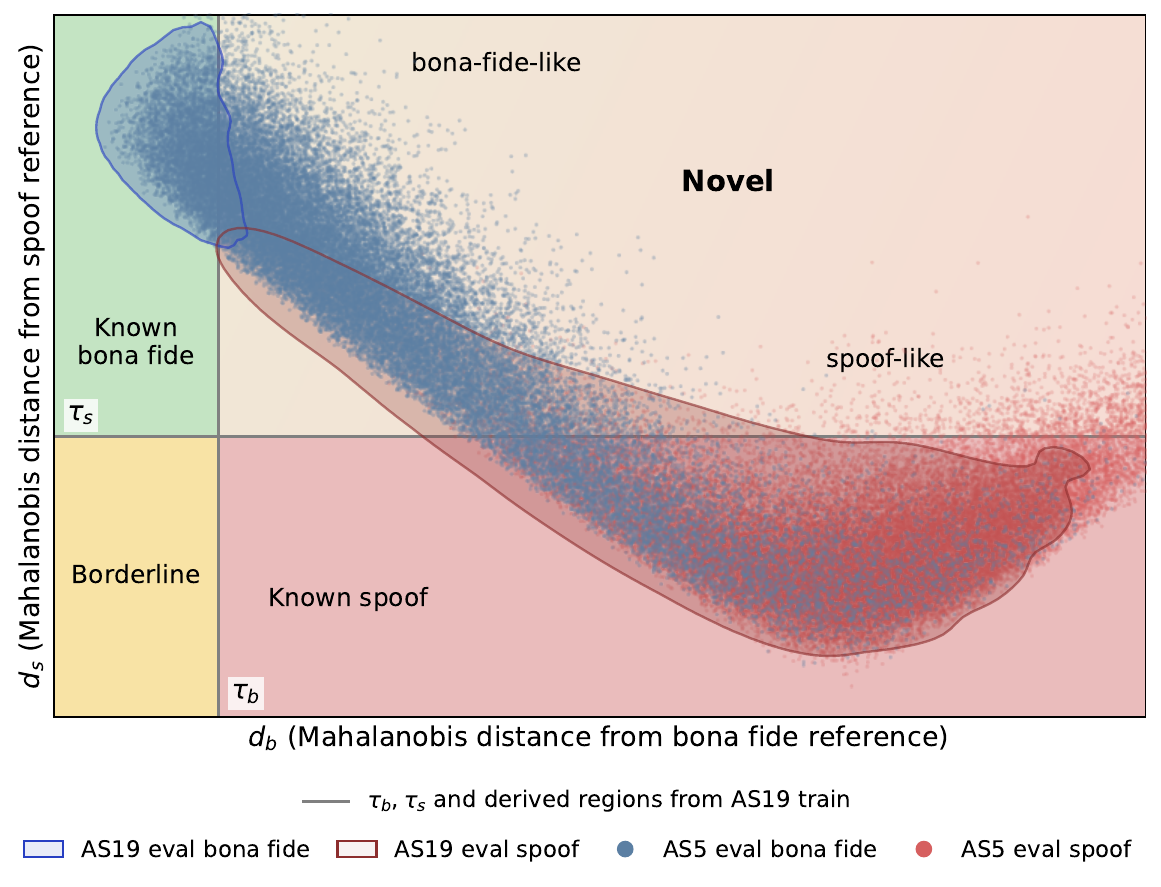}
\vspace{-2.2em}
\caption{Two-reference monitoring space for the MHFA detector trained on AS19. The colored regions and thresholds $\tau_b$ and $\tau_s$ are derived from AS19 train. Filled contours show AS19 eval distributions from the same benchmark, while points show AS5 eval samples from a different benchmark. AS19 eval spoof samples drift toward the bona fide side, while AS5 eval bona fide samples drift substantially across $\tau_b$ into spoof regions, illustrating the data drift captured by the monitor.}
\label{fig:distance_space}
\vspace{-1.3em}
\end{figure}

The pair $(d_b,d_s)$ defines a 2D decision space. The thresholds $\tau_b$ and $\tau_s$ split this space into four operating regions, as shown in \autoref{fig:distance_space}. Samples close only to the bona fide reference are treated as known bona fide, samples close only to the spoof reference are treated as known spoof, samples close to both references are treated as borderline, and samples far from both references are treated as novel, which can be further divided into bona fide/spoof-like novelty according to the closer reference, which provides a more detailed interpretation. 

This categorization is not supposed to replace the detector's binary classifier, but estimate whether the classifier is operating in a familiar part of its embedding space. The known bona fide and spoof regions correspond to familiar operating regimes. The novel region indicates that the recording is far from both references, suggesting an input drift. The borderline region is the most critical case: the embedding is simultaneously close to both the bona fide and spoof references, so the detector is operating in an ambiguous region of the representation space where automatic decisions should not be trusted.

\subsection{Operating policy and evaluation}

At deployment, the monitor is used as a gate for automatic processing. Samples assigned to the novel or borderline regions are alerted and routed to review. Samples assigned to known regions are processed automatically using the detector's binary prediction. Here, an alert does not imply a correct detector prediction. It means the system does not automatically accept unfamiliar or ambiguous input; it can be reviewed by a human, rejected, or handled in accordance with deployment policy.
For evaluation, we combine the detector decision and the monitor category into three operational outcomes. A sample is \emph{correctly trusted (CT)} when it is assigned to a known region, and the detector's binary decision is correct. A sample is \emph{alerted} when the monitor routes it to review (i.e., categories novel or borderline), regardless of the detector decision. A sample is \emph{wrongly trusted (WT)} when it is assigned to a known region, but the detector decision is wrong; this is a critical failure mode: the detector made an incorrect automatic decision, and the monitor did not flag the input as unreliable.

\subsection{Experiment setup}

We use the described monitoring procedure on four pretrained detectors: XLS-R-300M as the SSL extractor with either an MHFA or AASIST backend. Each detector is trained either on ASVspoof~2019~LA (AS19) or on ASVspoof~5 (AS5) and outputs a score in $[0,1]$, where higher values indicate the spoof class\footnote{Code for reproducing the monitor, detector details, and experiment scripts are available at \url{https://github.com/Security-FIT/Drift-Monitor}.}. 
We evaluate each detector on AS19 eval, AS5 eval, and In-the-Wild (ITW). The first two represent a (semi)-controlled benchmark conditions from different ASVspoof generations, while In-the-Wild represents a more realistic open-set setting. For a detector trained on AS19, AS19 data are considered in-domain, and all other datasets are out-of-domain. Analogously, AS5 is in-domain for detectors trained on AS5. This protocol tests whether the monitor captures operational mismatch relative to the detector's training reference. 
\autoref{tab:detector_eer} shows that each detector performs best in-domain and degrades on mismatched datasets, confirming the open-set reliability problem that motivates drift monitoring. 

\begin{table}[htbp]
\centering
\vspace{-1em}
\caption{Detector performance measured by EER~(\%) across evaluation datasets. The first column shows the training dataset of each detector. Light gray cells mark the in-domain performance.}
\label{tab:detector_eer}
\begin{tabular}{llrrr}
\toprule
Train & Detector & AS19 & AS5 & ITW \\
\midrule
\multirow{2}{*}{AS19}
    & AASIST & \cellcolor{gray!15}0.13 & 10.86 & 6.96 \\
    & MHFA   & \cellcolor{gray!15}1.37 & 17.47 & 7.31 \\
\midrule
\multirow{2}{*}{AS5}
    & AASIST & 7.84 & \cellcolor{gray!15}3.60 & 8.60 \\
    & MHFA   & 18.20 & \cellcolor{gray!15}4.36 & 18.18 \\
\bottomrule
\end{tabular}
\vspace{-1em}
\end{table}


For assessing the functionality of the proposed monitor, we report the three operational outcomes defined above. We fix the detector operating threshold at $0.5$ for comparability and additionally report deepfake false acceptance rate (FAR), i.e., the fraction of deepfake recordings classified as bona fide. After monitoring, deepfake FAR corresponds to the critical case where a deepfake recording is classified as bona fide and \textit{wrongly trusted}, i.e., accepted automatically without an alert.
Changing the classification threshold would change detector errors and therefore CT, WT, and FAR, but not the monitor categories or alert rates, which depend only on $d_b$, $d_s$, and the reference thresholds $\tau_b$, $\tau_s$.


\section{Drift Monitor Results}


Firstly, we evaluate whether the proposed monitor can detect the described loss of reliability from detector embeddings. \autoref{tab:monitor_results} reports the resulting operational outcomes. Unlike the detector alone, the monitor prevents unfamiliar inputs from being processed automatically. This changes the deployment behavior: many samples that would otherwise receive a detector decision are instead alerted. A major observation is that out-of-domain alert rates are often elevated, especially on ITW. This is the desired behavior of the monitor: when presented with new data, many embeddings no longer fall within the familiar regions defined by the training references. A high alert rate indicates that the underlying detector is poorly matched to the incoming distribution and may require adaptation or retraining. As a result, the monitor does not blindly trust the detector output, but exposes unreliable inputs before they can become silent errors.

Non-negligible alert rates also appear in some in-domain settings. This is expected to some extent as the ASVspoof training and evaluation splits are strictly disjoint across speakers and synthesizers~\cite{todisco2019asvspoof, wang24_asvspoof}. Thus, ``in-domain'' only means that the recordings come from the same benchmark, not necessarily that they follow the same distribution as the training reference. The monitor provides a granular insight: it asks whether a sample resembles the actual training reference used to define the trusted regions, rather than simply coming from the same benchmark.
The most pronounced case is AS5, where alert rates remain high even for detectors trained on AS5. We therefore treat this case separately. A closer inspection shows that the alerts are not caused only by spoof attacks, but also by the drift of bona fide recordings.
This is presumably related to the heavy codec and conditions variability introduced in AS5, as well as the fact that eval attacks are disjoint from the training ones. This suggests that AS5 evaluation data serve as a reasonable generalization test even for AS5-trained detectors, and that bona fide drift is an important factor in deployment.

Overall, the monitor behaves conservatively: it detects dataset mismatch and raises alerts under distribution shift, but the alert threshold is stricter than usual in-domain evaluation. This behavior is partly a design choice. We use the 95th percentile of training self-distances as a simple initial calibration rule, which makes the monitor more cautious and suitable for e.g., forensic triage, where an unreliable decision is more costly than sending additional samples for review. In other deployments, the threshold can be adjusted to achieve a more lenient trade-off between trusted and alerted cases.

\begin{table}[t]
\centering
\caption{Operational monitor outcomes of correctly trusted (CT), alerted, and wrongly trusted (WT) inputs. All values are percentages. Light-gray cells indicate the in-domain dataset for the corresponding detector.}
\label{tab:monitor_results}
\setlength{\tabcolsep}{3.2pt}
\begin{tabular}{@{}llccc|ccc|ccc@{}}
\toprule
\multirow{2}{*}{Train} & \multirow{2}{*}{Detector}
& \multicolumn{3}{c}{AS19}
& \multicolumn{3}{c}{AS5}
& \multicolumn{3}{c}{ITW} \\
\cmidrule(lr){3-5}
\cmidrule(lr){6-8}
\cmidrule(lr){9-11}
& & CT & Alert & WT
& CT & Alert & WT
& CT & Alert & WT \\
\midrule
\multirow{2}{*}{AS19}
& AASIST
& \cellcolor{gray!15}82.1 & \cellcolor{gray!15}17.9 & \cellcolor{gray!15}0.00
& 73.1 & 23.2 & 3.74
& 24.3 & 75.5 & 0.27 \\
& MHFA
& \cellcolor{gray!15}78.8 & \cellcolor{gray!15}21.0 & \cellcolor{gray!15}0.20
& 76.2 & 20.3 & 3.42
& 41.0 & 58.3 & 0.67 \\
\midrule
\multirow{2}{*}{AS5}
& AASIST
& 0.2 & 99.7 & 0.13
& \cellcolor{gray!15}22.0 & \cellcolor{gray!15}78.0 & \cellcolor{gray!15}0.01
& 9.3 & 90.7 & 0.01 \\
& MHFA
& 4.1 & 95.1 & 0.76
& \cellcolor{gray!15}18.9 & \cellcolor{gray!15}81.1 & \cellcolor{gray!15}0.00
& 20.5 & 79.5 & 0.01 \\
\bottomrule
\end{tabular}
\vspace{-2.4em}
\end{table}

\subsection{Preventing critical deepfake acceptance}

Secondly, we isolate the most security-critical error: deepfake recordings that are falsely accepted as bona fide (deepfake FAR). Without monitoring, these errors go unnoticed. With monitoring, they are dangerous only if the monitor also assigns them to a trusted region. \autoref{tab:s2b_before_after} therefore compares the deepfake FAR before and after utilizing the drift monitor.

\begin{table*}[htbp]
\centering
\caption{Deepfake false acceptance rate (FAR) before and after applying the monitor. Each Deepfake FAR cell reports the detector-only failure before monitoring, followed by the uncaught failure after monitoring. Reduction is the relative reduction in deepfake FAR, computed before rounding. Light-gray cells indicate the in-domain dataset.}
\label{tab:s2b_before_after}
\setlength{\tabcolsep}{4pt}
\renewcommand{\arraystretch}{1.15}
\begin{tabular}{ll|rr|rr|rr}
\toprule
\multirow{2}{*}{Train} & \multirow{2}{*}{Detector}
& \multicolumn{2}{c}{AS19}
& \multicolumn{2}{c}{AS5}
& \multicolumn{2}{c}{ITW} \\
\cmidrule(lr){3-4}
\cmidrule(lr){5-6}
\cmidrule(lr){7-8}
& & \multicolumn{1}{c}{Deepfake FAR} & Reduction
  & \multicolumn{1}{c}{Deepfake FAR} & Reduction
  & \multicolumn{1}{c}{Deepfake FAR} & Reduction \\
\midrule
\multirow{2}{*}{AS19}
& AASIST
& \cellcolor{gray!15}$3.20\% \rightarrow 0.00\%$ & \cellcolor{gray!15}\textbf{99.95\%}
& $0.02\% \rightarrow 0.00\%$ & \textbf{100.00\%}
& $0.88\% \rightarrow 0.00\%$ & \textbf{100.00\%} \\
& MHFA
& \cellcolor{gray!15}$5.49\% \rightarrow 0.22\%$ & \cellcolor{gray!15}\textbf{96.04\%}
& $0.05\% \rightarrow 0.00\%$ & \textbf{100.00\%}
& $1.51\% \rightarrow 0.00\%$ & \textbf{100.00\%} \\
\midrule
\multirow{2}{*}{AS5}
& AASIST
& $59.43\% \rightarrow 0.15\%$ & \textbf{99.75\%}
& \cellcolor{gray!15}$32.28\% \rightarrow 0.02\%$ & \cellcolor{gray!15}\textbf{99.94\%}
& $55.01\% \rightarrow 0.02\%$ & \textbf{99.96\%} \\
& MHFA
& $54.08\% \rightarrow 0.85\%$ & \textbf{98.43\%}
& \cellcolor{gray!15}$17.29\% \rightarrow 0.00\%$ & \cellcolor{gray!15}\textbf{100.00\%}
& $61.72\% \rightarrow 0.02\%$ & \textbf{99.97\%} \\
\bottomrule
\end{tabular}
\vspace{-1.7em}
\end{table*}

The monitor significantly reduces the most dangerous failure mode. Before monitoring, several detectors silently accept a large fraction of deepfake recordings as bona fide, especially for detectors trained on AS5. For example, the AS5-trained AASIST detector accepted 59.43\% of AS19 deepfakes as bona fide, and the AS5-trained MHFA detector accepted 61.72\% of ITW deepfakes. After applying the monitor, the uncaught deepfake FAR drops below 1\% across all settings. 
The relative reduction exceeds 98\% across all AS5 detectors, preventing the false acceptance of the vast majority of deepfakes.

For AS19-trained detectors, the deepfake FAR is already much lower, but the same trend holds. This does not mean that these detectors make few errors overall. Rather, a larger share of their errors is concentrated in the other direction: bona fide recordings being rejected as deepfake. Such errors are still operationally undesirable, as they may increase false alarms or unnecessary manual review, but they do not indicate that the system silently accepts a deepfake recording as bona fide. This is why we isolate deepfake FAR as the critical security case.


Importantly, these improvements do not mean that the detector itself becomes more accurate. The monitor does not correct detector decisions; instead, it identifies embeddings that fall outside trusted regions and prevents them from being automatically accepted. Thus, many errors that would otherwise go unnoticed are converted into system alerts. The reduction in uncaught deepfake FAR is therefore achieved by reducing automatic coverage, as reflected by high alert rates in \autoref{tab:monitor_results}. \textbf{The fundamental security gain is obtained by making detector uncertainty observable through the drift monitor, allowing uncertain samples to be routed to review or rejected.} This is the intended operating mode for high-risk applications: the monitor trades automatic coverage for a lower probability of accepting deepfake recordings as bona fide.

\subsection{Effect of covariance regularization}
\label{sec:regularization}

We ablate covariance regularization because Mahalanobis distances depend on the inverse reference covariance. When detector embeddings are highly anisotropic, empirical covariance estimates can be poorly conditioned, leading to low-variance directions dominating the distance and resulting in unstable or ``alert everything'' behavior. We compare the empirical covariance $\Sigma$ with the shrinkage covariance $\widehat{\Sigma}$ used in the main results, without changing the detector or its binary decision threshold, as described in \autoref{sec:method}.

\begin{figure*}[t]
\centering
\includegraphics[width=0.75\textwidth]{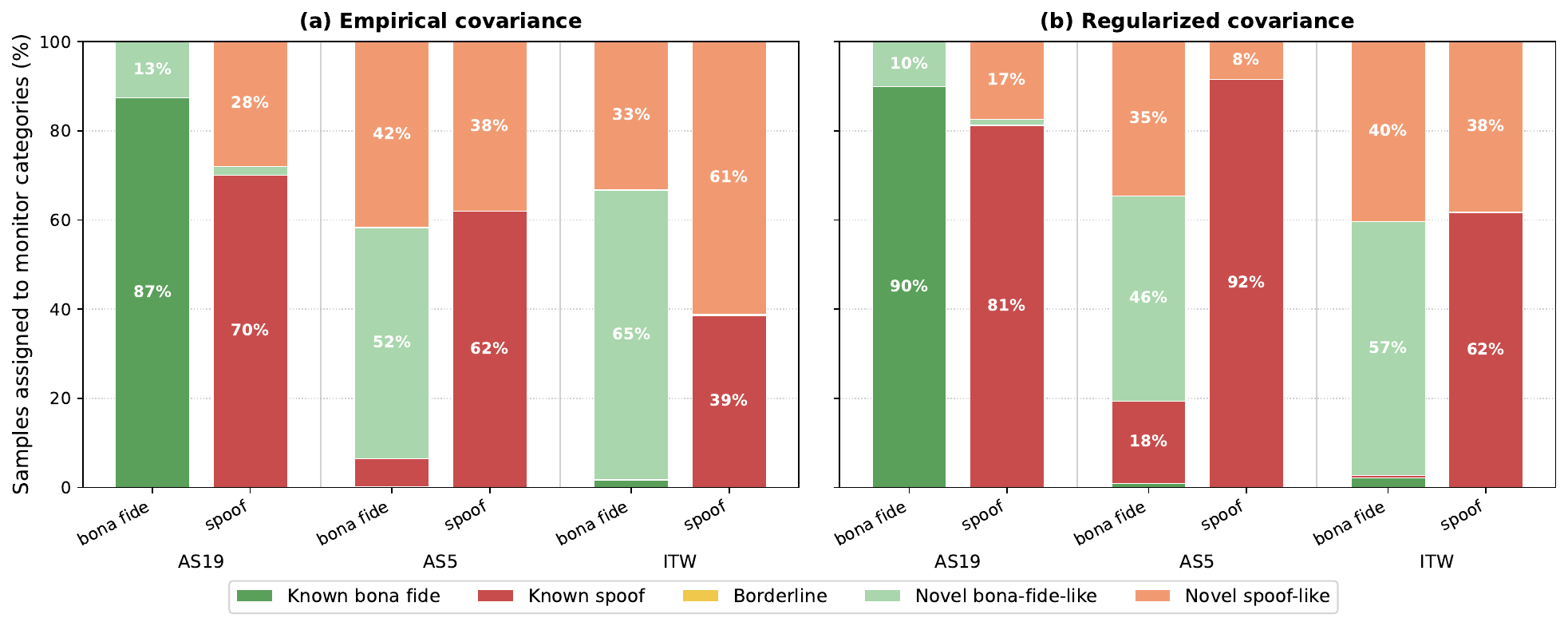}
\vspace{-1em}
\caption{Category assignment before and after covariance regularization for AASIST trained on AS19. Bars are normalized within each dataset and class. The bars show how the monitor categorized a specific evaluation subset (per class), whereas \autoref{tab:cov_reg_ablation} reports aggregate alert and FAR rates (per detector).}
\label{fig:cov_reg_categories}
\vspace{-1em}
\end{figure*}

\autoref{tab:cov_reg_ablation} shows that regularization changes the trusted-region geometry in a backend-dependent way. For AASIST, shrinkage reduces apparent over-alerting, especially for the AS19-trained detector, while high alert rates remain for AS5-trained detectors, indicating genuine dataset novelty rather than only a covariance artifact. For MHFA, regularization often increases alert rates but reduces uncaught deepfake FAR in most settings. Thus, shrinkage is not simply stricter or more lenient; it reshapes the category distribution, reduces artificially inflated distances, and trades geometric specificity for stable Mahalanobis estimates.
\autoref{fig:cov_reg_categories} illustrates this effect for AASIST trained on AS19. 
Regularization moves many samples toward known regions, while substantial novelty persists in mismatched datasets. This illustrates that covariance regularization reduces artificial alerts without eliminating genuine dataset drift.
The AS5 spoof column also illustrates the trade-off introduced by regularization. Although more bona fide data were confidently assigned to the spoof category for AS5, the fraction of AS5 spoof samples automatically assigned to the known-spoof region increases from 62\% to 92\%. Thus, regularization substantially increases automatic handling at the cost of a small increase in bona fide rejection for the AS19-trained AASIST.

\begin{table}[ht]
\centering
\vspace{-1em}
\caption{Effect of covariance regularization on monitor alert rate and deepfake FAR. Empirical covariance $\mathbf{\Sigma}$ is compared with the regularized covariance $\mathbf{\widehat{\Sigma}}$ used in the main results. All values are percentages.}
\label{tab:cov_reg_ablation}
\setlength{\tabcolsep}{5pt}
\renewcommand{\arraystretch}{1.05}
\begin{tabular}{@{}llr|rr|rr|rr}
\toprule
\multirow{2}{*}{Train} & \multirow{2}{*}{Detector} & \multirow{2}{*}{Metric}
& \multicolumn{2}{c}{AS19}
& \multicolumn{2}{c}{AS5}
& \multicolumn{2}{c}{ITW} \\
\cmidrule(lr){4-5}
\cmidrule(lr){6-7}
\cmidrule(lr){8-9}
& & & \multicolumn{1}{c}{$\mathbf{\Sigma}$} 
    & \multicolumn{1}{c}{$\mathbf{\widehat{\Sigma}}$} \vline
    & \multicolumn{1}{c}{$\mathbf{\Sigma}$} 
    & \multicolumn{1}{c}{$\mathbf{\widehat{\Sigma}}$} \vline
    & \multicolumn{1}{c}{$\mathbf{\Sigma}$} 
    & \multicolumn{1}{c}{$\mathbf{\widehat{\Sigma}}$} \\
\midrule
\multirow{4}{*}{AS19}
& \multirow{2}{*}{AASIST}
& Alert
& 28.1 & 17.9
& 49.3 & 23.2
& 84.6 & 75.5 \\
& & FAR
& 0.00 & \cellcolor{gray!20}0.00
& 0.00 & \cellcolor{gray!20}0.00
& 0.00 & \cellcolor{gray!20}0.00 \\
\cmidrule(lr){2-9}
& \multirow{2}{*}{MHFA}
& Alert
& 9.4 & 21.0
& 17.7 & 20.3
& 45.2 & 58.3 \\
& & FAR
& 0.94 & \cellcolor{green!20}0.22
& 0.00 & \cellcolor{gray!20}0.00
& 0.00 & \cellcolor{gray!20}0.00 \\
\midrule
\multirow{4}{*}{AS5}
& \multirow{2}{*}{AASIST}
& Alert
& 100.0 & 99.7
& 95.3 & 78.0
& 96.4 & 90.7 \\
& & FAR
& 0.00 & \cellcolor{red!15}0.15
& 0.01 & \cellcolor{red!15}0.02
& 0.01 & \cellcolor{red!15}0.02 \\
\cmidrule(lr){2-9}
& \multirow{2}{*}{MHFA}
& Alert
& 81.5 & 95.1
& 67.7 & 81.1
& 78.5 & 79.5 \\
& & FAR
& 1.49 & \cellcolor{green!20}0.85
& 0.01 & \cellcolor{green!20}0.00
& 0.09 & \cellcolor{green!20}0.02 \\
\bottomrule
\end{tabular}
\flushleft
\scriptsize
\emph{Note:} Only deepfake FAR changes are color-coded, because lower FAR is preferable, whereas lower or higher alert rates depend on the trade-off between automatic coverage and monitor sensitivity. Green indicates a FAR decrease after regularization, red indicates a FAR increase, and gray indicates no change after rounding.
\vspace{-1.7em}
\end{table}

\section{Discussion and Conclusion}

In this work, we treat deepfake speech detection as a deployment reliability problem, not only as a binary classification task. The proposed unsupervised drift monitor uses internal detector embeddings to estimate whether an input lies within a familiar region defined by bona fide and spoof training references. The monitor therefore extends the detector and monitors whether the detector output can be trusted automatically. 

The main result is that the drift monitor substantially reduces the silent acceptance of deepfakes. Many deepfake recordings that would be falsely accepted as bona fide by the detector are instead alerted by the monitor and sent for additional review. This is especially relevant for forensic and other high-risk settings, where missing a deepfake is more costly than increasing manual review. The alert rate, however, depends on threshold calibration: the 95th-percentile rule used here is a conservative initial operating point and should be adjusted to the desired trade-off between automatic coverage and monitor sensitivity.

The results also show that drift affects both classes. Deepfake attacks can move toward the bona fide reference, but bona fide recordings can also drift because of codec, channel, or recording-condition changes. This explains high alert rates even in some in-domain settings. Deployment monitoring should therefore track both bona fide and deepfake drift and act accordingly.
Finally, covariance regularization is necessary for stable Mahalanobis monitoring because detector embeddings can be highly anisotropic. Shrinkage trades some geometric specificity for numerically stable distance estimates, making the monitor usable across different backends.

Importantly, this work focuses on drift detection and monitoring. The monitor can identify samples that should not be trusted automatically, but it does not decide how the detector should be updated, how unknown data should be labeled, or how retraining should be performed. These mitigation strategies, including active learning, reference updating, and continual detector adaptation, are an emerging research domain~\cite{wang2022dataselection, furuhashi2024dataselection} and a natural direction for future work. Also, the monitor's behavior depends on the selected threshold: the 95th percentile rule provides an initial setup, but different applications may require different trade-offs. Importantly, the monitor depends mainly on the quality and geometry of the detector embeddings and on the representativeness of the training references. Finally, our experiments are limited to offline benchmark evaluation, and future work should study online monitoring with human-review workflows.

Our proposed monitor provides an unsupervised mechanism for deciding when a deployed speech deepfake detector should be trusted, alerted, or treated as operating outside its familiar domain. To the best of our knowledge, this is among the first structured methods for monitoring data drift in deployed detectors, and it represents an important step toward reliable deepfake detection systems for real-world, security-critical scenarios.

\bibliographystyle{IEEEtran}
\bibliography{IEEEabrv, bibliography}

\end{document}